\documentclass[sigplan,nonacm]{acmart}

\usepackage{fancyhdr}
\usepackage{times}  
\usepackage{titlesec}
\usepackage{tikz}
\usepackage{amsmath}
\usepackage{graphicx}
\usepackage{xspace}
\usepackage{xcolor}
\usepackage[T1]{fontenc}
\usepackage{subfigure}
\usepackage{verbatim}
\usepackage{url}
\usepackage{enumitem}
\usepackage{listings}
\usepackage{multirow}
\usepackage[ruled,linesnumbered]{algorithm2e}
\usepackage{pifont}
\usepackage{mathtools}
\usepackage{fancyhdr}
\usepackage{caption}
\usepackage{xurl}

\author{Yongchao He}
\affiliation{
  \institution{ScitiX AI}
  \city{~}
  \country{~}
  }

\author{Bohan Zhao}
\affiliation{
  \institution{ScitiX AI}
  \city{~}
  \country{~}
  }

\author{Zheng Cao}
\affiliation{
  \institution{ScitiX AI}
  \city{~}
  \country{~}
  }
  
\newcommand{\sysname}{SiPipe\xspace} %
\newcommand{\sysnames}{SiPipe’s\xspace} %

\newcommand{\num}[1]{\normalsize{\textcircled{\scriptsize{#1}}}\normalsize\xspace}

\newcommand{\crunch}{\vspace{-1mm}}

\newcommand{\circledzero}{%
  \tikz[baseline={([yshift=-0.3ex]char.base)},inner sep=0pt,minimum size=1.6ex]{
    \node[shape=circle, fill=black, text=white, font=\scriptsize] (char) {0};
  }%
}

\begin{document}
\title{\sysname: Bridging the CPU–GPU Utilization Gap for Efficient Pipeline-Parallel LLM Inference}

\begin{abstract}
As inference workloads for large language models (LLMs) scale to meet growing user demand, pipeline parallelism (PP) has become a widely adopted strategy for multi-GPU deployment, particularly in cross-node setups, to improve key-value (KV) cache capacity and inference throughput. However, PP suffers from inherent inefficiencies caused by three types of execution bubbles—\emph{load-imbalance}, \emph{intra-stage}, and \emph{inter-stage}—which limit pipeline saturation.
We present \sysname, a heterogeneous pipeline design that improves throughput by leveraging underutilized CPU resources to offload auxiliary computation and communication. \sysname incorporates three key techniques—\emph{CPU sampling}, a \emph{token-safe execution model}, and \emph{structure-aware transmission}—to mitigate pipeline bubbles and improve execution efficiency.
Across diverse LLMs, \sysname achieves up to 2.1$\times$ higher throughput, 42.7\% lower per-token latency, and up to 23\% higher average GPU utilization compared to the state-of-the-art vLLM under the same PP configuration, demonstrating its generality across LLMs and deployment scenarios.
\end{abstract}

\maketitle 


\section{Introduction}



Large language models (LLMs), e.g., GPT~\cite{ouyang2022gpt, brown2020gpt3} and Llama~\cite{llama2followup, dubey2024llama3.1}, have achieved remarkable success in inference tasks such as code generation and question answering, enabling widely used applications like ChatGPT~\cite{chatgpt}. However, this success has come with rapidly increasing model sizes—often reaching hundreds of billions of parameters—which far exceed the memory capacity of a single GPU—and with it, a growing user base that demands ever higher inference throughput~\cite{infer-trend}.

Scaling LLMs across multiple GPUs is essential due to memory limits~\cite{agrawal2024taming}. For example, while a 72B LLM’s weights may fit on two 80GB GPUs, the key-value (KV)~\cite{qin2025mooncake} caches used to accelerate inference typically require 2–3$\times$ more memory, increasing the practical GPU requirement to eight or more~\cite{llama2followup, vllm_tunning, shoeybi2019megatron}. To handle such large-scale deployments, \emph{tensor parallelism (TP)} is commonly employed to distribute computation across GPUs. However, TP’s reliance on frequent \emph{all-reduce}~\cite{patarasuk2009allreduce} communications creates a performance bottleneck beyond 4–8 GPUs, as inter-GPU communication during \emph{forward} passes often dominates computation time~\cite{shoeybi2019megatron}.

\begin{figure}[t]
    \centering
    \setlength{\abovecaptionskip}{-2pt}
    \includegraphics[width=\linewidth]{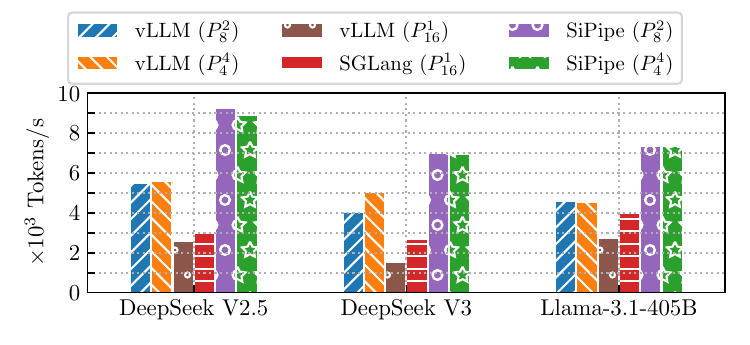}
    \caption{Throughput comparison of different engines under various parallel configurations on 16 H100 GPUs. Each configuration is denoted as $P^{i}_{j}$, where the PP degree $p = i$ and the TP degree $t = j$. See \S\ref{sec:exp_setup} for detailed experimental settings.}
    \label{fig:bench_tput_16GPU_H100}
    \crunch
    \crunch
    \crunch
\end{figure}

To scale LLM inference, \emph{pipeline parallelism (PP)} complements TP by dividing the model into sequential stages, each mapped to one or more GPUs. Within each stage, TP~\footnote{Data parallelism is treated as TP degree $t=1$.} is used to further parallelize the \emph{forward} of that stage—i.e., a portion of the model—across the assigned GPUs. This hybrid strategy reduces inter-GPU communication and enables efficient scaling across large GPU clusters. As discussed in prior works~\cite{shoeybi2019megatron, zhong2024distserve, fu2024serverlessllm, 2024Sarathi}, let $p$ and $t$ denote the degrees of PP and TP, respectively, and let $N = p \times t$ be the total number of GPUs used. The overall inference throughput scales as $T(p, t) \propto \frac{1}{\frac{k_1}{pt} + \frac{k_2 \log t}{p} + b}$, while the per-token latency follows $D(p, t) \propto \frac{k_1}{t} + k_2 \log t + k_3 (p - 1)$, where $k_i$ and $b$ are constants. A detailed derivation is provided in Appendix~\ref{appendix}.
Therefore, given a user-facing inference service with a latency budget $D_e$ (i.e., a service-level objective, or SLO), the inference engine must select an empirical configuration $(p_e, t_e)$ such that $D(p_e, t_e) \le D_e$, subject to the constraint $p_e \times t_e = N$, while maximizing throughput. This trade-off typically requires a carefully tuned PP degree $p_e$—large enough to exploit parallelism, yet small enough to satisfy the latency target.
Figure~\ref{fig:bench_tput_16GPU_H100} validates this conclusion. On a 16-GPU, 2-node deployment, recent LLMs achieve peak throughput with PP degrees $p = 2$ or $4$, using state-of-the-art (SOTA) inference engine vLLM~\cite{kwon2023vllm}. For example, on DeepSeek V3~\cite{liu2024deepseekv3}, the PP-based configuration vLLM ($P^{4}_{4}$) delivers up to 3.22$\times$ throughput of the pure TP configuration vLLM ($P^{1}_{16}$).

Despite recent progress, mainstream inference engines~\cite{kwon2023vllm, zheng2024sglang, DeepSpeed} offer limited support for PP, missing opportunities to scale LLM inference efficiently across multiple GPUs. Furthermore, their naïve PP designs~\cite{kwon2023vllm, DeepSpeed, shoeybi2019megatron} introduce several fundamental bottlenecks (\S\ref{sec:observation}):


\noindent
\textbf{Stage-wise compute load imbalance.}
PP introduces a compute imbalance absent in pure TP: although LLM layers are evenly divided across stages, the last stage incurs extra compute overhead from the \emph{sampling} task (\S\ref{subsec:autoregressive}), resulting in uneven stage durations and pipeline bubbles. 

\noindent
\textbf{Stage-wise input preparation overhead.}  
Before each \emph{forward} pass, the CPU prepares input tensors and transfers them to GPU memory, incurring a preparation overhead \(t_p\), while the \emph{forward} itself costs \(t_f\). In PP, this preparation is redundantly repeated at every stage, even though each stage computes only a \(1/p\) fraction of the \emph{forward}. As a result, the preparation-to-computation ratio increases from \(\frac{t_p}{t_f}\) to \(\frac{t_p}{t_f/p}\), leading to greater GPU idle time as the PP degree \(p\) increases.

\noindent
\textbf{Unpredictable cross-stage communication overhead.} 
In PP, strict data dependencies between adjacent stages—where the size of the transferred data vary with each iteration—require synchronized communication. This communication pattern exhibits unique characteristics of \emph{communication stalls} and \emph{multi-round metadata exchanges}, resulting in overheads that far exceed the cost of the data transfer itself.

Collectively, these factors highlight the growing imperative for pipeline-aware architecture in LLM serving systems. Although recent studies explore diverse pipelined strategies \cite{ma2024hpipe, wang2024pipefusion, tan2025pipellm, butler2024pipeinfer}, they leave core pipeline bottlenecks unresolved. Likewise, specialized designs for serverless inference~\cite{guo2025gllm} or Mixture-of-Expert (MoE) models \cite{zhu2025megascaleinfer, chen2023pipeline_moe} address narrow scenarios but fall short of providing a general-purpose, inference-oriented pipeline framework.


To tackle these bottlenecks, we identify a key inefficiency in LLM inference: while GPU computation is almost saturated and KV caching with inter-node communication imposes heavy memory and network pressure, CPUs on the same nodes are often underutilized (typically below 10\%, see \S\ref{sec:utilization}). \sysname exploits this \emph{CPU slack} by offloading parts of inter-stage communication and auxiliary tasks to underutilized CPUs, mitigating pipeline stalls and improving throughput. This insight motivates the design of \sysname, a \emph{\underline{\textbf{S}}aturated \underline{\textbf{I}}nference \underline{\textbf{Pipe}}line} that rebalances computation across CPUs and GPUs for more efficient PP. 

\noindent
\textbf{\sysname decouples \emph{sampling} from GPU execution through \emph{asynchronous, column-wise CPU sampling}.}
\sysname performs sampling on the CPU to mitigate GPU-side computation imbalance across stages. It adopts a column-wise layout for tensors and performs incremental updates, enabling in-place computation with minimal memory allocation. This design reduces CPU latency and eliminates sampling-induced stalls in PP, thereby improving overall inference efficiency. (\S\ref{sec:cpu_sampling})

\noindent
\textbf{\sysname decouples and overlaps CPU and GPU execution with the \emph{token-safe execution model (TSEM)}.}  
\sysname introduces the TSEM to overlap CPU-side input preparation with GPU-side \emph{forward}. For a given batch size, TSEM pre-captures two versions of the CUDA graph, which alternately bind to two shared buffers. This design allows the CPU to fill one buffer while the GPU reads the other, enabling seamless parallelism without modifying the execution graph. By decoupling input preparation from \emph{forward}, TSEM eliminates intra-stage GPU stalls and improves overall throughput. (\S\ref{sec:token-safe-exec})

\noindent
\textbf{\sysname introduces \emph{structure-aware transmission (SAT)} for efficient and decoupled stage communication.}  
Based on the insight that \emph{hidden states} (i.e., the output tensors of each stage) exhibit structural stability across inference iterations, SAT infers tensor layouts in advance, enabling early memory allocation and eliminating the need to transmit metadata between stages. This significantly reduces communication rounds and removes the overhead of metadata serialization and deserialization. By allowing receivers to predict tensor layouts ahead of time, SAT eliminates synchronous metadata exchange, naturally enabling asynchronous communication and mitigating \emph{communication stalls}. (\S\ref{sec:structure-aware-trans})




Our evaluation demonstrates that under the same PP configurations, \sysname achieves a throughput improvement of 1.6$\times$-2.1$\times$ and 1.4$\times$-1.7$\times$ over vLLM~\cite{kwon2023vllm} on the $8 \times$H100 and $16 \times$H100 testbeds, respectively, along with latency reductions of up to 30.5\% and 42.7\%.  
As shown in Figure~\ref{fig:bench_tput_16GPU_H100}, \sysname delivers up to 4.5$\times$ throughput improvement over the pure TP approach in a cross-node setup with 16 H100 GPUs.

In summary, the contributions of this paper are:  

\noindent
(1) We identify three types of PP bubbles caused by pipeline-agnostic designs in existing LLM inference engines that significantly degrade performance. (\S\ref{sec:observation})

\noindent
(2) We analyze PP characteristics and propose leveraging underutilized CPU resources to assist communication and computation, mitigating these bubbles. (\S\ref{sec:challenges})

\noindent
(3) We design and implement \sysname based on these insights; the code will be open-sourced after publication. (\S\ref{sec:design})

\noindent
(4) We conduct extensive experiments on representative LLMs, demonstrating that \sysname outperforms SOTA systems such as vLLM~\cite{kwon2023vllm} and SGLang~\cite{zheng2024sglang}. (\S\ref{sec:evaluation})


\section{Background}
\label{sec:background}

\subsection{LLM Inference}
\label{subsec:autoregressive}
\begin{figure}[t]
    \centering
    \includegraphics[width=\columnwidth]{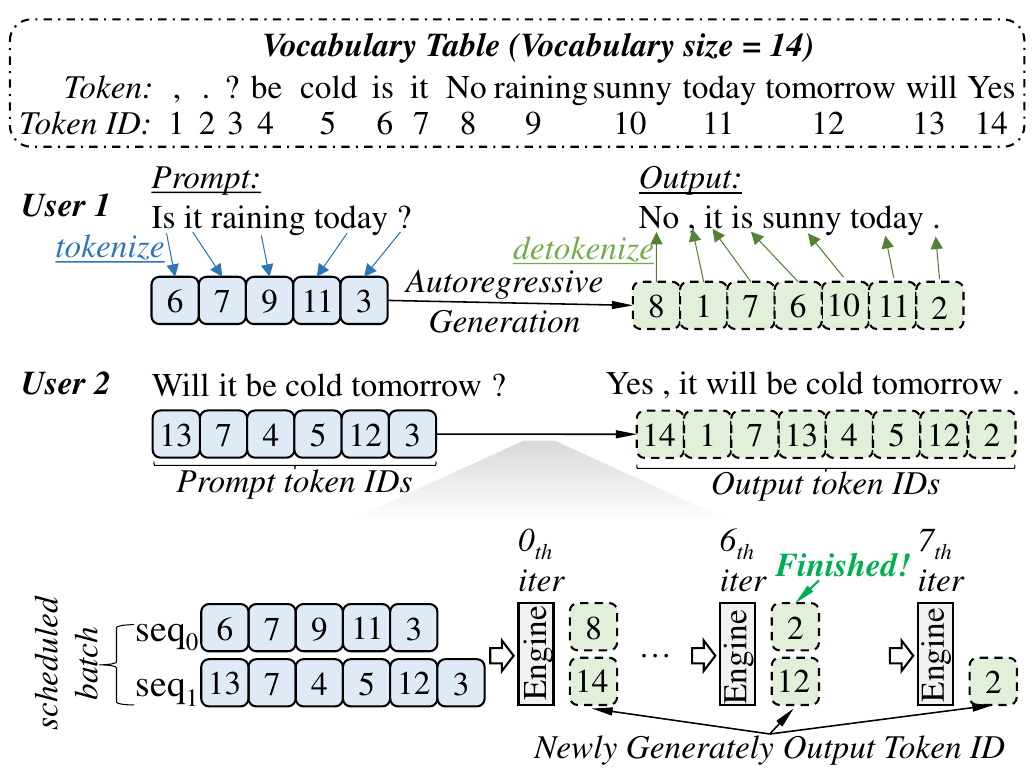}
    \caption{Example of iterative autoregressive generation in LLM serving. Note that the scheduled batch may changed from one iteration to the next—for instance, certain sequences may be preempted by others. The diagram omits these dynamics for clarity.}
    \label{fig:autoregressive}
    \crunch
    \crunch
    \crunch
\end{figure}

Figure~\ref{fig:autoregressive} illustrates the \emph{autoregressive generation} process of LLM inference, where a pretrained LLM (e.g., DeepSeek-R1~\cite{guo2025deepseek}) generates \textbf{output} tokens conditioned on a \textbf{prompt}. A tokenizer~\cite{gage1994new, sennrich-etal-2016-neural, kudo-richardson-2018-sentencepiece} \emph{tokenizes} the prompt into discrete \textbf{token IDs} using a fixed \emph{vocabulary table}, which maintains the mapping between IDs and tokens.
An inference \emph{engine}~\cite{zheng2024sglang, kwon2023vllm} maintains the generation state for each request as a \textbf{sequence}, which consists of the token IDs of both the input prompt and all output token IDs. Generation proceeds in discrete \textbf{iterations}, each producing one new token ID per scheduled sequence. To improve hardware efficiency, sequences at the same iteration are grouped into a \textbf{batch} for parallel inference on GPUs.
Each decoding iteration consists of two parts\footnote{We focus on the \emph{decoding} stage in this paper, where token generation proceeds one step at a time, in contrast to the initial \emph{prefill}~\cite{qin2025mooncake} stage that encodes the full prompt in a single pass.}:

Outside the stages, the engine performs two key tasks. First, it selects a batch of active sequences (e.g., $seq_0$ and $seq_1$ at iteration 0) and generates a \emph{scheduling output}, which is broadcast to all GPUs. Completed sequences are removed, and new ones are added to maintain batch occupancy and reduce queuing latency.  
Second, after the batch finishes traversing all stages, the engine appends the generated token IDs to each sequence, checks for completion, and \emph{detokenizes} the outputs into human-readable text.

Inside each stage, three steps are applied in every iteration.

\noindent
\emph{\num{1} Input preparation.}
Each GPU constructs a \emph{model input} using the received \emph{scheduling output}. This step encompasses tasks such as model and attention metadata computation, sequence caching and updates, tensor allocation, and CPU–GPU data transfers. The resulting model input is a collection of tensors that are subsequently transferred from CPU to GPU.

\noindent
\emph{\num{2} Forward pass.}
In PP, each stage processes model inputs and returns an \emph{activation} vector for each sequence. Non-final stages produce intermediate activations of shape $B \times H$, where $B$ is the microbatch size and $H$ the hidden dimension, and forward them to the next stage. Only the final stage transforms these into \emph{logits} of shape $B \times V$, where $V$ is the vocabulary size. Each row of the logits corresponds to a sequence, and each column to a vocabulary token.


\noindent
\emph{\num{3} Sampling.}
A \emph{softmax} in the final stage converts logits into probabilities. 
We denote the \emph{logits} from the final layer of a LLM at time step \(s\) as $\mathbf{z}_s \in \mathbb{R}^{B \times V}$,
where \(B\) is the batch size and \(V\) is the vocabulary size. For each request \(b \in \{1, \dots, B\}\), \(\mathbf{z}_s^{(b)} \in \mathbb{R}^V\) represents the unnormalized log-probabilities over the vocabulary.
Let \(\mathbf{y}_{<s}^{(b)} = (y_1^{(b)}, \dots, y_{s-1}^{(b)})\) denote the sequence of previously generated tokens for request \(b\) up to step \(s{-}1\).  
The sampling process can be defined as follows.

\noindent
\emph{(1) Logits adjustment.} Modify logits using penalties~\cite{penalty-fre-pre, penalty-rep} (e.g., frequency) based on \(\mathbf{y}_{<s}^{(b)}\) to promote diversity.

\noindent
\emph{(2) Probability computation.} Scale logits by temperature \(\tau\)~\cite{temperature-scaling}, apply \emph{softmax}, and optionally restrict candidates using top-\(k\)~\cite{top-k} or top-\(p\)~\cite{top-p} filtering:
$$
\mathbf{p}_i^{(b)} = \mathrm{Filter}\left(
  \mathrm{softmax}\left(
    \frac{
      \mathrm{ApplyPenalty}(\mathbf{z}_s^{(b)}, \mathbf{y}_{<s}^{(b)})
    }{\tau}
  \right);\, k, p
\right)
$$

\noindent
\emph{(3) Sampling.} Draw a token ID from the categorical distribution $y_s^{(b)} \sim \mathrm{Categorical}(\mathbf{p}_s^{(b)}), \quad \mathbf{y}_s \in \mathbb{N}^B$.

For example, at the $0_\text{th}$ iteration, top-$k$ sampling may select token IDs $8$ and $14$ for $seq_{0}$ and $seq_{1}$ respectively, yielding updated sequences $seq_{0} = \langle 6, 7, 9, 11, 3, \underline{8} \rangle$ and $seq_{1} = \langle 13, 7, 4, 5, 12, 3, \underline{14} \rangle$.
The engine proceeds to the next iteration, repeating until an \textsc{End-of-Sequence (EOS)} token ID is generated or the maximum sequence length is reached.



\subsection{Pipeline Parallelism in LLM Inference}
LLMs typically consist of stacked transformer~\cite{vaswani2017transformer} layers, each performing attention and feed-forward computations. Pipeline parallelism (PP) partitions the model along its depth, with each \emph{stage} responsible for a subset of layers.
During inference, unfinished sequences are grouped into \emph{microbatches} and processed in a pipelined fashion. For example, with four sequences $\{seq_0, seq_1, seq_2, seq_3\}$ and pipeline degree $p=2$, one schedule assigns microbatch $\{seq_0, seq_1\}$ to iteration $2k$ and $\{seq_2, seq_3\}$ to $2k+1$, alternating until completion. Each stage sequentially performs \emph{input preparation} and a \emph{forward pass}, while only the final stage executes \emph{sampling}. 

PP enables inference of LLMs that do not fit on a single GPU and improves throughput. Unlike TP, whose \emph{allreduce} operations scale poorly across nodes, PP restricts communication to adjacent stages and is more bandwidth-efficient —especially in cross-node deployments using Ethernet or InfiniBand.
Let $N$, $H$, $B$, $t$, and $p$ denote the number of layers, hidden size, batch size, TP degree, and PP degree, respectively. The per-GPU communication volume is $C(p, t) = BH\left(\frac{4N(t-1)}{pt} + p - 1\right)$. When $p^2t < 4N$—a condition commonly satisfied in LLM deployments that require 8 or more GPUs under PP (e.g., $N > 50$, $pt \leq 16$)~\cite{qwen2.5, liu2024deepseekv3, llama2followup}—we have $C(p, t) > 2 \times C(2p, t/2)$, suggesting that increasing $p$ and reducing $t$ can substantially lower communication.
For small models\footnote{Not in the scope of this paper.} (e.g., 32B LLMs that require $pt \leq 4$), PP brings limited benefit and may introduce pipeline bubbles. But for larger models ($pt \geq 8$), hybrid PP+TP (e.g., $p=2$, $t=4$) significantly reduces communication versus pure TP ($p=1$, $t=8$). As shown in Figure~\ref{fig:bench_tput_16GPU_H100}, PP yields up to 4.5$\times$ throughput improvement for 16-GPU inference of DeepSeek V3.

\section{Motivation}
\label{sec:motivation}
This section presents key observations on PP and identifies core challenges in achieving high throughput. These observations reveal fundamental inefficiencies in PP inference and motivate the design of \sysname. 

\subsection{Observations of PP Inefficiencies in LLM Inference}
\label{sec:observation}

\begin{figure}[t]
    \centering
    \setlength{\abovecaptionskip}{-1pt}
    \includegraphics[width=\linewidth]{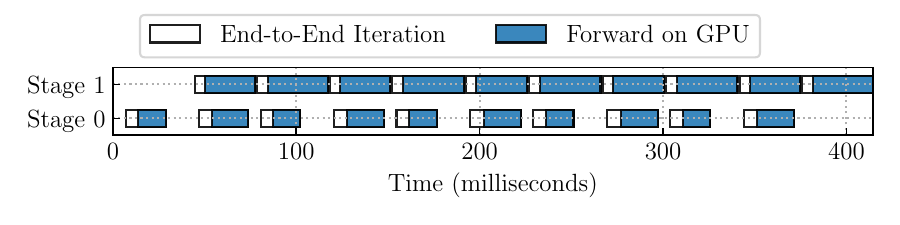}
    \caption{Per-iteration execution breakdown using vLLM on 8 H100 GPUs with Qwen-2.5-72B ($t=2$, $p=4$). Bars denote \emph{iteration time}; filled regions represent GPU \emph{forward time}.}
    \label{fig:breakdown_iteration_qwen_72B}
    \crunch
    \crunch
    \crunch
\end{figure}

\begin{figure}[t]
    \centering
    \setlength{\abovecaptionskip}{-1pt}
    \includegraphics[width=\linewidth]{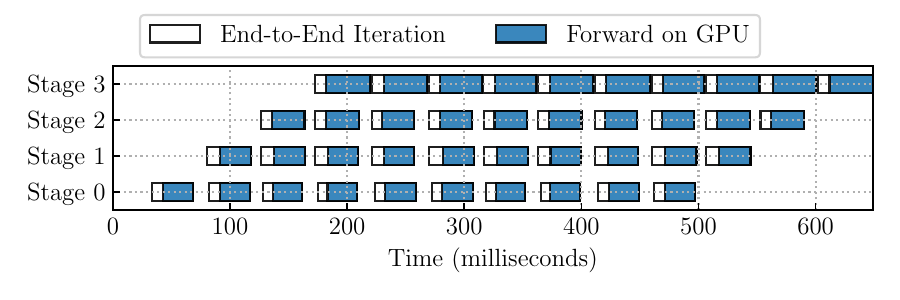}
    \caption{Per-iteration execution breakdown using vLLM on 16 H100 GPUs with DeepSeek V3 ($t=4$, $p=4$).}
    \label{fig:breakdown_iteration_deepseek_v3}
    \crunch
    \crunch
    \crunch
\end{figure}

Figure~\ref{fig:breakdown_iteration_qwen_72B} and Figure~\ref{fig:breakdown_iteration_deepseek_v3} show the per-stage breakdown of each iteration during 8-GPU and 16-GPU PP inference with vLLM~\cite{kwon2023vllm}. These results lead to the following observations:

\noindent
\textbf{Observation 1: Load-imbalance bubble.} The overall throughput of PP inference is limited by the slowest stage. Profiling across various LLMs and GPU counts consistently shows the final stage has a significantly higher computation load ($22\%-40\%$)—mainly due to the extra cost of \emph{sampling}. This imbalance causes earlier stages to idle: with PP degree $p = x$, stage~0 at iteration~$n+x$ must wait for stage~$x-1$ to complete iteration~$n$. Consequently, the overloaded final stage creates persistent pipeline bubbles, reducing overall efficiency.

\noindent
\textbf{Observation 2: Intra-stage bubble.}
In each iteration, we observe a gap ($12\%-19\%$) at the beginning of every stage's execution, caused by the CPU-side \emph{input preparation} before launching the GPU-side \emph{forward}. This delay arises from the use of CUDA graphs~\cite{guide2020cuda}, which require inputs to reside at fixed GPU memory addresses. As a result, CPU-prepared tensors must be synchronously copied into pre-allocated buffers before each CUDA graph replay. 

In frameworks like vLLM, input preparation is deferred until the previous \emph{forward} completes to ensure correctness, avoiding race conditions in graph execution. This forces the GPU to wait for the CPU to finish preparation, creating intra-stage bubbles. While more asynchronous designs are possible, they risk correctness issues (see Section~\ref{sec:challenges}).

\noindent
\textbf{Observation 3: Inter-stage bubble.}
\emph{inter-stage bubbles} become non-negligible after an even $p$-stage partition, which consist of two components: synchronization overhead due to data dependencies of adjacent stages ($1.4\text{--}2.6$ ms) and  transmission overhead of activations ($1\text{--}2$ ms). Figures~\ref{fig:breakdown_iteration_qwen_72B} and~\ref{fig:breakdown_iteration_deepseek_v3} reveal that even with an even $p$-stage partition, the \emph{forward} execution time varies across stages with a standard deviation of $3\%-7\%$. This variation amplifies synchronization cost.

For example, suppose stage 0 takes $t^{n+1}_0$ to complete iteration $n+1$, while stage 1 takes $t^n_1$ to complete iteration $n$, where $t^{n+1}_0 < t^n_1$. If both begin at time $t$, stage 0 becomes idle since $t + t^{n+1}_0$, waiting to send hidden states to stage 1. The interval $[t + t^{n+1}_0,\ t + t^n_1]$ is a  \emph{communication stall} induced solely by stage-wise synchronization on data dependencies.

\subsection{Challenges and Solutions}
\label{sec:challenges}

Eliminating bubbles in PP requires addressing challenges in computation efficiency, dependency management, and overlapping communication. We next detail each challenge and show how underutilized CPUs are leveraged to mitigate them.

\noindent
\textbf{Challenge 1: \emph{Decoupling sampling to preserve pipeline balance.}}  
To eliminate load-imbalance bubbles, we first identify their root cause: \emph{The final pipeline stage performs an additional sampling step}. 
In typical PP, sampling is naturally colocated with the final stage, as this stage produces the logits. However, this design introduces imbalance: while other stages proceed to the next microbatch after \emph{forward}, the final stage is stalled by the additional, sequential sampling step. This creates a serialization point that degrades overall throughput.

A natural idea is to offload sampling to a dedicated GPU. However, in large-scale deployments where all GPUs are fully utilized (e.g., $t=2$, $p=4$ on an 8-GPU node), these options are impractical. Sampling is too light and sequential to warrant an exclusive GPU, and reshaping the pipeline to hide its latency (such as giving the final stage fewer layers) forces hardware-specific workload balancing, reducing portability. Because these strategies entangle sampling with model partitioning, they also impede modular scaling. Efficiently decoupling sampling thus remains a key challenge.

\noindent
\textbf{Solution 1: \emph{\sysname adopts a column-wise layout and incremental computation to accelerate CPU-based sampling.}}
\sysname offloads the sampling process to multiple CPUs to balance GPU workload in PP. To avoid the CPU becoming a straggler during high-throughput LLM inference, \sysname introduces a column-wise data layout to enable efficient in-place updates. Built on this layout, \sysname incrementally updates cross-iteration metadata, avoiding redundant computation and memory allocation. These optimizations exploit inter-batch similarity, reduce CPU latency, and eliminate sampling-induced load-imbalance bubbles. (\S\ref{sec:cpu_sampling})

\noindent
\textbf{Challenge 2: \emph{Cross-iteration input conflicts under static execution models.}}
A natural way to eliminate intra-stage bubbles is to decouple CPU and GPU execution—preparing inputs asynchronously on the CPU while the GPU performs the \emph{forward} pass. 
However, to minimize runtime overhead, modern inference engines often adopt static execution models, i.e., CUDA graphs~\cite{guide2020cuda}, which fix input bindings and kernel sequences. While efficient, this approach leads to subtle \emph{write-after-read (WAR) hazards}: when the CPU and GPU run asynchronously, inputs for iteration $j$ may overwrite buffers that the GPUs is still reading from iteration $i < j$. These conflicts arise from temporal asymmetry between CPU and GPU, rigid memory bindings in CUDA graphs, and the evolving nature of model inputs in autoregressive generation.

A common workaround—staging model input tensors in temporary buffers and copying them into static regions before launch \emph{forward}—avoids conflicts but incurs extra memory copies, reintroducing synchronization bottlenecks and offsetting the benefits of CUDA Graphs.
These limitations call for a principled solution that enables safe decoupling without sacrificing the efficiency of static execution.

\noindent
\textbf{Solution 2: \emph{\sysname adopts token-safe execution model (TSEM) to decouple CPU and GPU execution.}}
To eliminate intra-stage bubbles caused by GPU stalls during CPU-side input preparation, \sysname introduces the \emph{TSEM}, which safely decouples input preparation from \emph{forward} while preserving static graph constraints. TSEM pre-captures two CUDA graphs per batch size, each bound to a distinct shared buffer. Iterations alternate between these graphs, allowing the CPU to prepare inputs in one buffer while the GPU reads the other. This \emph{shadow buffering}-based scheme~\cite{he2023generic} ensures version isolation without requiring runtime graph modifications. By enabling asynchronous CPU-GPU execution, TSEM eliminates intra-stage bubbles. (\S\ref{sec:token-safe-exec})

\noindent
\textbf{Challenge 3: \emph{Efficient transmission of dynamic hidden states across pipeline stages.}}
In PP, each stage transmits a tensor dictionary of \emph{hidden states} (activations and residuals~\cite{he2016resnet}) to the next stage every iteration. Since tensor sizes vary dynamically and cannot be statically pre-allocated, packing and transmission must wait until \emph{forward} passes complete, delaying communication. This uncertainty causes \emph{communication stall} (\S\ref{sec:observation}) and requires \emph{multi-round metadata exchanges} to determine tensor sizes, inflating inter-stage bubbles and reducing pipeline efficiency.

\noindent
\textbf{Solution 3: \emph{\sysname adopts a structure-aware transmission (SAT) strategy to enable asynchronous communication between stages.}}
Although hidden states change across iterations, their structure remains mostly stable. \sysname exploits this property via SAT, which captures the invariant structure and infers per-iteration changes from scheduling output—eliminating the need to treat each tensor dictionary as fully dynamic. Once the structure is identified, an asynchronous CPU thread in the receiving process handles data pre-allocation, management, and transmission for each iteration. As a result, downstream stages can directly access the prepared data without waiting, eliminating transmission latency. SAT reduces multi-round metadata exchanges, avoids redundant metadata processing, shortens inter-stage bubbles, and removes communication stalls through its asynchronous design. (\S\ref{sec:structure-aware-trans})
\begin{figure}[t]
    \centering
    \setlength{\abovecaptionskip}{-1pt}
    \includegraphics[width=\linewidth]{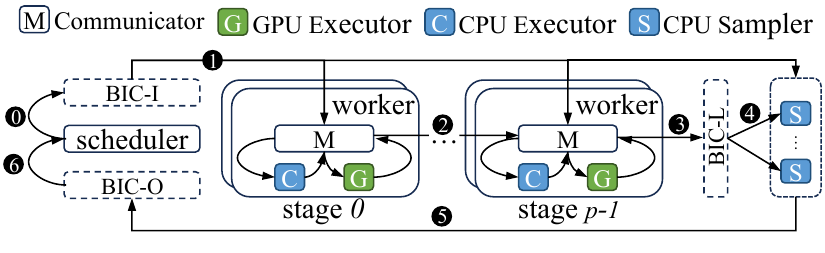}
    \caption{Architecture and workflow of \sysname.}
    \label{fig:overview}
    \crunch
    \crunch
    \crunch
    \crunch
    \crunch
    \crunch
\end{figure}

\section{Overview}
\label{sec:overview}


\subsection{System Architecture}
Figure~\ref{fig:overview} shows the pipelined parallel inference architecture of \sysname, comprising four key components: the \emph{scheduler}, \emph{workers}, \emph{CPU samplers}, and the \emph{Buffered IPC Channel (BIC)}.
In each iteration, the scheduler selects a microbatch from unfinished sequences and generates a \emph{scheduling output}, which is broadcast to all workers. The pipeline consists of $p$ stages, each with $t$ \emph{workers} executing TP inference over a model shard. The output of stage $i$ (i.e., the hidden states) becomes the input to stage $i{+}1$.
Each worker includes three modules:
(1) a \emph{communicator} that handles inter-stage data transfer and internal coordination;
(2) a \emph{CPU executor} for input preparation and cached state management;
(3) a \emph{GPU executor} for the \emph{forward} pass.
After the final stage, a pool of \emph{CPU samplers} receives the logits and performs parallel sampling.
To enable low-latency communication between components, \sysname introduces the BIC (detailed in \S\ref{sec:implementation}), which serves as a message queue for different data types: \emph{BIC-I} for scheduling outputs, \emph{BIC-L} for logits, and \emph{BIC-O} for sampling outputs. 

\subsection{Execution Flow}

We illustrate the \sysname workflow with an example. Suppose we process 4 sequences ($\{seq_0, seq_1, seq_2, seq_3\}$) with a PP degree of $p{=}2$, resulting in a microbatch size of $\frac{4}{p} = 2$.
At initialization, the scheduler dispatches $p$ consecutive scheduling outputs to workers via \emph{BIC-I} (\circledzero). Thereafter, upon receiving sampling output for iteration $n$, it immediately dispatches iteration $n+p$. This pipelined approach keeps all workers busy, maximizing resource utilization. For example, the scheduler may assign microbatches $\{seq_0, seq_1\}$ and $\{seq_2, seq_3\}$ to iteration $2k$ and iteration $2k+1$, respectively.

Each worker handles sequences as follows: the \emph{CPU executor} asynchronously updates the sequence cache, generates model input, and forwards it to the \emph{communicator}. Concurrently, the \emph{GPU executor} polls the communicator and launches \emph{forward} passes upon receiving new model input.
For stage-0 workers, execution triggers on model input availability alone (\ding{182}). Subsequent stages wait for both model input and hidden states from the previous stage. Intermediate stages forward hidden states downstream (\ding{183}), while the final stage sends logits to CPU samplers via \emph{BIC-L} (\ding{184}\ding{185}).

Each CPU sampler then generates sampling metadata for the microbatch and performs sampling using the logits. Sampling outputs are sent back to the scheduler via \emph{BIC-O} (\ding{186}), triggering the next iteration (\ding{187}).

\section{Design}
\label{sec:design}

In this section, we present \sysnames core designs for eliminating pipeline inefficiencies, focusing on load-imbalance bubbles (\S\ref{sec:cpu_sampling}), intra-stage bubbles (\S\ref{sec:token-safe-exec}), and inter-stage bubbles (\S\ref{sec:structure-aware-trans}).

\subsection{Column-Wise CPU Sampling via Metadata Reuse}
\label{sec:cpu_sampling}
Offloading sampling to the CPU reduces load-imbalance bubbles by balancing GPU workloads but introduces two main challenges. First, sampling relies on dynamic metadata (e.g., output token IDs) that change each iteration. Intermediate steps—such as penalty application, softmax, and filtering—produce many temporary tensors, causing frequent memory allocations and stressing CPU memory management. Second, sampling is compute-intensive and requires careful CPU-side optimization to maintain high throughput.

For instance, given logits \(\mathbf{Z} \in \mathbb{R}^{B \times V}\) and previously output token IDs \(\mathbf{Y} \in \mathbb{N}^{B \times (s-1)}\), we compute a penalty tensor \(\mathbf{f} = \mathrm{CalcPenalty}(\mathbf{Y})\), where each row \(\mathbf{f}_{i,:}\) encodes token frequency, presence, or repetition in sequence \(i\). The penalized logits are \(\mathbf{Z}' = \mathbf{Z} - \alpha \cdot \mathbf{f}\), with \(\alpha\) as a tunable hyperparameter. Then temperature scaling is applied: \(\mathbf{Z}'' = \mathbf{Z}' / \tau\), \(\tau > 0\), followed by a row-wise softmax \(\mathbf{P}_{i,j} = \frac{\exp(\mathbf{Z}''_{i,j})}{\sum_{v=1}^V \exp(\mathbf{Z}''_{i,v})}\). Filtering methods such as top-\(k\) and top-\(p\) sampling optionally restrict candidate tokens before sampling. Since \(\mathbf{Y}\) changes every iteration, these computations and memory allocations must be repeated.


Dynamic \(\mathbf{Y}\) and \(\mathbf{Z}\) impose significant overhead when batch size \(B\) and vocabulary size \(V\) are large—a common setting in LLM serving. For instance, Qwen-2.5-72B (\(V=151{,}643\), \(B=256\)) requires approximately 300\,MB per penalty tensor (double precision). To sustain high inference throughput, sampling must finish within the decoding slack—typically 1–2\,ms per sampling strategy—demanding over 150\,GB/s memory bandwidth solely for penalty operations. While this falls below DDR5’s peak bandwidth, contention under high concurrency reduces effective utilization. Additionally, element-wise computations on large matrices further increase computational cost as \(B\) and \(V\) grow. Without memory reuse or incremental updates, penalty calculation becomes a critical CPU bottleneck, limiting overall throughput.


To mitigate memory allocation overhead and facilitate reuse across iterations, we propose a \emph{column-wise data layout} that transposes the model logits \(\mathbf{Z} \in \mathbb{R}^{B \times V}\) and \(\mathbf{Y} \in \mathbb{N}^{B \times (s-1)}\) into \(\mathbf{Z}^\top \in \mathbb{R}^{V \times B}\) and \(\mathbf{Y}^\top \in \mathbb{N}^{(s-1) \times B}\), respectively. This reordering forms the basis of a redesigned CPU sampling that directly improves both \emph{computational efficiency} and \emph{memory access locality}.
Specifically, \sysname reuses this transposed structure to incrementally construct penalties and optimize downstream sampling steps:

\noindent
\emph{(1) Memory reuse for output sequences.} We preallocate a maximum-length output buffer \(\mathbf{Y} \in \mathbb{N}^{L_{\text{max}} \times B}\). In the transposed layout, new token IDs are appended as rows, enabling fast in-place update and avoiding tensor reshaping. This also ensures that all subsequent metadata (e.g., frequency matrices) can be updated incrementally.

\noindent
\emph{(2) Incremental penalty construction.} For penalties such as frequency, repetition, and presence, we maintain a shared buffer\footnote{Each penalty type has its own buffer and follows a similar processing routine.} \(\mathbf{f} \in \mathbb{R}^{V \times B}\) aligned with the transposed logits \(\mathbf{Z}^\top\). At each iteration, only the \(B\) elements of \(\mathbf{f}\) that correspond to the newly generated token IDs are incrementally updated in-place across all sequences in the microbatch. This approach enables efficient computation of \(\alpha \cdot \mathbf{f}\) with minimal overhead, while being cache-friendly and avoiding reconstruction of the entire penalty tensor. Consequently, penalty adjustment reduces to a single vectorized subtraction $\mathbf{Z}^\top := \mathbf{Z}^\top - \alpha \cdot \mathbf{f}$, eliminating the need to regenerate \(\mathbf{f}\) from scratch at every iteration.


\noindent
\emph{(3) Low-cost pipelined integration.} In each stage, each worker typically generates a logits shard of shape \(B \times \frac{V}{t}\). By employing a column-wise layout, these shards can be locally transposed to \(\frac{V}{t} \times B\) and directly concatenated along the row dimension to form the global matrix \(\mathbf{Z}^\top \in \mathbb{R}^{V \times B}\). This approach avoids costly all-gather operations or row-wise tensor reconstructions.

By redesigning the sampling process to leverage the column-major layout, we unify key sampling operations into in-place transformations on the logits tensor \(\mathbf{Z}^\top\). This eliminates repeated intermediate tensor allocations and data copying. All operations run in a memory- and cache-efficient manner, enabling vectorized computation and high throughput. The resulting design forms a lean, scalable CPU-side sampling pipeline that is both elegant and efficient in practice.

\sysname relies on the assumption that consecutively scheduled batches are identical or exhibit high similarity (i.e., the microbatch size $B$ remains constant across iterations), thereby permitting incremental updates through minor adjustments to \(\mathbf{Y}\). Although this assumption may appear restrictive, it is well-justified in large-scale LLM inference scenarios employing scheduling strategies such as \emph{contiguous batching} or \emph{iteration batching}~\cite{kwon2023vllm, agrawal2024taming, fu2024serverlessllm, yu2022orca}. In particular, within TP, adjacent batches are highly likely to be identical. Similarly, in PP with parallelism degree \(p\), batches indexed by \(n\) and \(n+p\) are expected to be identical, while batches indexed by \(n\) and \(n+j\) for \(0 < j < p\) are orthogonal. Consequently, maintaining only \(p\) distinct column-wise data replicas—corresponding to the pipeline stages—is sufficient to support efficient incremental computation without violating this assumption.

The memory overhead on the host for each replica is bounded by \(n \times B \times V + B \times L_{\max}\), where \(B\) typically does not exceed 1024, and \(n=3\) corresponds to the frequency, presence, and repetition penalty buffers. Moreover, \(L_{\max}\) and $V$ for most SOTA LLMs rarely exceed 128K and 200K, respectively. Consequently, the total memory consumption remains well within reasonable limits ($\approx $ 3-4 GB), making this overhead acceptable for typical host memory capacities.

\subsection{Token-Safe Execution Model with CPU-Assist}
\label{sec:token-safe-exec}

To eliminate intra-stage bubbles, \sysname decouples CPU and GPU executors for asynchronous execution. In PP, microbatches scheduled across successive iterations are non-overlapping sequence sets. Therefore, the CPU executor can safely prepare inputs for iteration \(i+1\) while the GPU executor performs \emph{forward} on iteration \(i\). Without data dependencies between adjacent microbatches, CPU-GPU concurrency is both safe and effective. This insight underpins our decoupled execution strategy, enabling full utilization of CPU and GPU resources without compromising correctness.

However, this asynchronous CPU-GPU execution conflicts with CUDA graphs, which rely on static buffer bindings that assume inputs remain unchanged across iterations. 
While CUDA graphs reduce launch overhead by capturing fixed kernel execution sequences and their memory bindings, this static model causes WAR hazards when the CPU overwrites data still in use by the GPU, threatening correctness (see Challenge 2, \S\ref{sec:challenges}).

\begin{figure}[t]
    \centering
    \setlength{\abovecaptionskip}{-1pt}
    \includegraphics[width=0.95\linewidth]{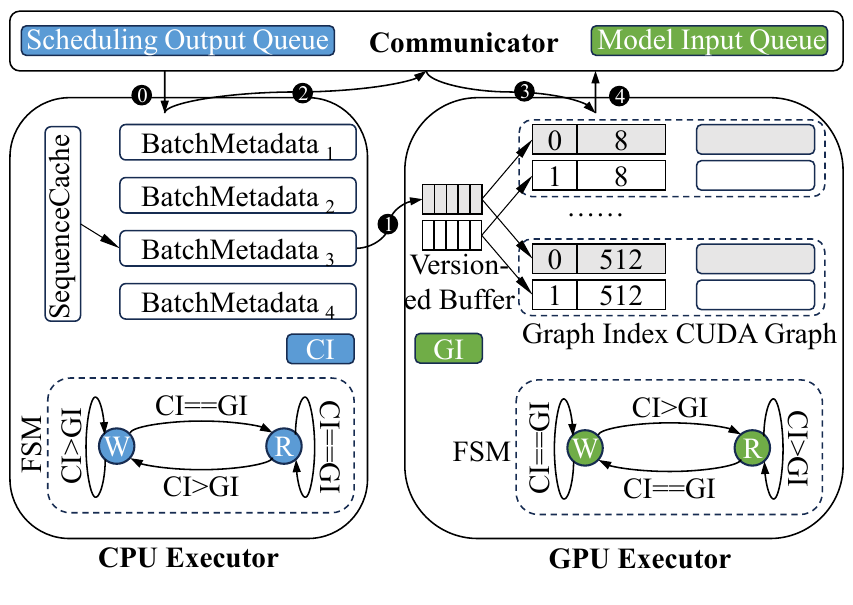}
    \caption{Illustration of the token-safe execution model for conflict-free GPU execution.}
    \label{fig:tsem}
    \crunch
    \crunch
    \crunch
\end{figure}

To address cross-iteration WAR hazards under static execution models, we propose the \emph{token-safe execution model (TSEM)}, a mechanism that ensures conflict-free pipeline inference. TSEM introduces an asynchronous CPU executor dedicated to input preparation, which operates in parallel with the GPU executor while maintaining compatibility with statically captured CUDA graphs. To coordinate these components without introducing blocking or direct dependencies, TSEM adopts a message-driven architecture (Figure~\ref{fig:tsem}), where a dedicated \emph{communicator} orchestrates execution by dispatching events to the CPU and GPU executors. This design decouples input preparation from \emph{forward} and eliminates idle time due to executor synchronization, allowing both components to make forward progress independently.

The GPU executor employs a versioned buffer mechanism, where each type of input tensor maintains two physical versions ($v_{0}$ and $v_1$). For each version, we capture a set of static CUDA graphs, each corresponding to a different batch size. A graph is selected based on a tuple \texttt{<$v_0$/$v_{1}$, $batch\_size$>}, enabling the executor to alternate between versions across iterations and avoid WAR hazards. The CPU executor maintains a \emph{SequenceCache} that maps each sequence to its cached metadata, including prompt/output token IDs and KV-related state, minimizing redundant data transfer when sequences are repeatedly scheduled. It also prepares per-iteration \emph{BatchMetadata}, which includes preprocessed CPU tensors (e.g., attention inputs) for each microbatch. To reduce recomputation, $p$ versions of BatchMetadata are maintained (where $p$ is PP degree), leveraging the high similarity between microbatches in iteration $i$ and $i{+}p$ (\S\ref{sec:cpu_sampling}). The communicator coordinates execution by managing two message queues: one for scheduling outputs and one for model inputs. It buffers scheduling outputs from the scheduler and model inputs from the CPU executor, and asynchronously transmits GPU executor outputs to the next stage while receiving hidden states from the previous one. To drive progress, it maintains two indicators—the CPU indicator (CI) and GPU indicator (GI)—which track the latest completed iteration of each executor.

Each executor operates as a finite-state machine (FSM) with two states: \texttt{wait (W)} and \texttt{running (R)}. Initially, both CI and GI are set to $-1$. The CPU executor transitions from W to R—or remains in R—whenever \texttt{CI == GI}, indicating that all previously generated model inputs have been consumed. In state R, it consumes a scheduling output from the communicator's \emph{scheduling output queue} (\circledzero), updates the SequenceCache and the corresponding BatchMetadata (indexed by $i \bmod p$, where $i = \texttt{CI}$), and writes the prepared input tensors into the versioned input buffer (version $i \bmod 2$, \ding{182}). It then enqueues a lightweight model input descriptor (e.g., batch size) into the model input queue and increments CI (\ding{183}). The GPU executor follows a similar FSM structure, with a key distinction: it increments GI immediately after entering or remaining in state R, thereby allowing the CPU executor to asynchronously prepare the next iteration’s input. Upon entering R, it fetches a model input from the model input queue (\ding{184}), locates the correct versioned buffer and CUDA graph using the index \texttt{<(GI-1)$\bmod$2, batch\_size>}, executes the graph, and forwards the output to the next stage via the communicator (\ding{185}). This FSM-driven design decouples CPU and GPU progress while maintaining strict correctness, enabling efficient overlap of input preparation and \emph{forward} pass.

TSEM eliminates intra-stage bubbles by decoupling CPU input preparation from GPU execution, ensuring that each new iteration can immediately begin once the previous \emph{forward} pass completes. This tight coordination maximizes pipeline utilization and sustains high throughput.
The overhead of TSEM is modest, requiring only an additional static input buffer and a few CUDA graphs.
This cost is effectively neutralized by offloading sampling to the CPU, which reclaims GPU memory previously allocated for sampling metadata, e.g., penalty tensor $\mathbf{f}$.

\subsection{Structure-Aware Hidden State Transmission}
\label{sec:structure-aware-trans}

In PP, consecutive stages exhibit strict data dependencies: the output (i.e., hidden states) of stage $i$ serves as the input to stage $i{+}1$ for every microbatch. This mandates a data handoff between adjacent stages at each iteration. 
Cross-stage communication is inherently serialized. Stage $i{+}1$ must wait until it fully receives and deserializes the hidden states from stage $i$ before initiating \emph{forward} pass. This strict dependency leads to \emph{inter-stage bubbles}, where communication latency hinders overall pipeline utilization. Since tensor shapes and sizes of hidden states may vary across microbatches, the receiving stage cannot initiate partial \emph{forward} and must wait for full transmission and deserialization.

\begin{figure}[t]
    \centering
    \setlength{\abovecaptionskip}{-1pt}
    \includegraphics[width=\linewidth]{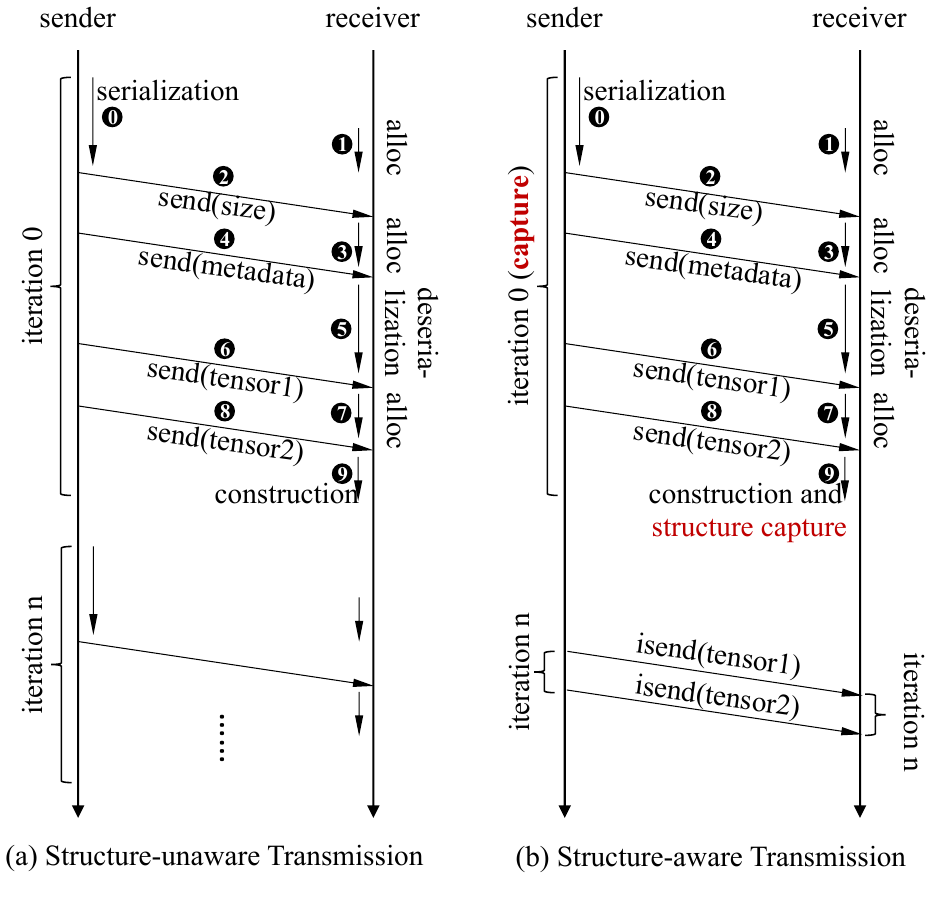}
    \caption{Effect of structure-aware vs. structure-unaware transmission when transferring variable-size tensors across multiple iterations.}
    \label{fig:struct-aware}
    \crunch
    \crunch
    \crunch
\end{figure}

Consider transmitting hidden states structured as \{$\mathsf{key1}$: $\mathsf{tensor}_1$, $\mathsf{key2}$: $\mathsf{tensor}_2$\}. As shown in Figure~\ref{fig:struct-aware}(a), state-of-the-art inference engines typically adopt a \emph{structure-unaware transmission} strategy. In this approach, the receiver lacks prior knowledge of the dictionary layout or tensor sizes, preventing asynchronous communication and enforcing strict synchronization.
The sender first serializes metadata containing the number of entries, the list of keys, and per-tensor attributes such as shape, dtype, and device (\circledzero). Because the receiver does not know the size of this metadata blob, it must first allocate a temporary buffer to receive the blob size (\ding{182}), then perform a communication round to obtain it (\ding{183}). Once the size is known, it allocates a full metadata buffer (\ding{184}) and receives the actual blob in a second communication round (\ding{185}). After deserialization (\ding{186}), the receiver sequentially allocates memory and receives each tensor one by one (\ding{187}–\ding{189}). 
Finally, it reconstructs the hidden state by pairing each key with its corresponding tensor (\ding{190}). This serialized, step-wise process requires multiple tightly coupled communication rounds and memory allocations, significantly delaying stage activation and exacerbating inter-stage bubbles.

For a running engine with the fixed model type and GPU assignment, the structure of the hidden state dictionary remains largely static across iterations. In typical models, this dictionary contains a small, fixed set of tensors, such as the output of the final feed-forward layer and the residual connection. As a result, both the number of entries and their string keys are predictable.
Each tensor typically has a shape of $x \times y$, where $y$ (the hidden size) is model-dependent and fixed, whereas $x$ (the batch size) may vary across iterations. All other attributes, including data type and device, remain unchanged.

Figure~\ref{fig:struct-aware}(b) illustrates \sysname's structure-aware transmission mechanism, which avoids redundant metadata communication by separating static and dynamic components of the hidden states. The mechanism consists of two key components: static structure capture and dynamic batch size retrieval.
In the first iteration, the receiver performs a full transmission using the structure-unaware transmission. During the reconstruction step, however, it additionally extracts and stores the static structure of the dictionary (\ding{190}).
Since the scheduler sends per-iteration scheduling output to each worker, the receiver can directly extract the only dynamic factor—the batch size—based on the number of sequences assigned to that stage.

With both static structure and batch size available, the receiver can pre-allocate memory and post asynchronous \texttt{irecv} operations before transmission begins, avoiding \emph{communication stalls} caused by unfinished \emph{forward} on GPU. The sender transmits hidden states using asynchronous \texttt{isend}, overlapping communication with ongoing computation. As shown in Figure~\ref{fig:struct-aware}(b), metadata-related serialization, deserialization, and synchronization are skipped entirely during steady-state iterations.

Our structure-aware mechanism provides two key advantages over the structure-unaware baseline. First, by removing synchronous metadata communication, it decouples adjacent pipeline stages. In the structure-unaware setting, stage $i$ may stall due to blocking communication. In contrast, our design enables pre-buffering and asynchronous receives to eliminate these communication stalls.
Second, although hidden states are small—typically requiring only hundreds of microseconds to transfer over RDMA or NVLink—the overhead from metadata handling and control synchronization in the structure-unaware design dominates communication cost. By bypassing these steps, our approach brings inter-stage latency down from $2\text{--}5$ ms to the pure data transfer overhead of less than $500\mu$s, which is further hidden by overlapping communication with computation asynchronously (reducing the effective inter-stage bubble to $\sim$$100\mu$s).


\section{Implementation}
\label{sec:implementation}

\sysname is implemented in pure Python to maximize compatibility with existing inference frameworks. Specifically, it is implemented as a plugin for \texttt{vLLM}~\cite{kwon2023vllm}, inheriting from key Python classes and overriding selected methods to integrate seamlessly. 

Unlike conventional frameworks where most computation is offloaded to GPUs, \sysname performs sampling on the CPU. Due to Python’s Global Interpreter Lock (GIL), we adopt a multi-process architecture to fully exploit CPU parallelism.\footnote{Future versions will explore Python 3.13’s \emph{free-threaded} mode to simplify the implementation.} This leads to substantial inter-process communication, which we categorize into two patterns: \emph{dispatch} and \emph{combine}.

For dispatch—broadcasting scheduling outputs from the scheduler to all worker and sampler processes, and broadcasting logits from the final-stage GPUs to all samplers—we design \emph{Buffered IPC Channels (BIC)} using shared-memory ring buffers with a \emph{lock-ahead} mechanism. We instantiate \emph{BIC-I} for scheduling output and \emph{BIC-L} for logits transfer. Each BIC maintains $N$ slots, each guarded by a mutex implemented using \texttt{fcntl} file locks, enabling lightweight, cross-process synchronization without busy-waiting. 
In iteration $n$, the producer pre-acquires an exclusive (write) lock on slot $(n+1) \bmod N$, writes data to slot $n \bmod N$, and then releases the $(n \bmod N)_{th}$ slot's exclusive lock—ensuring lock-ahead progression without contention.  
Consumers poll slots sequentially, acquiring a shared (read) lock when a slot becomes available, which allows concurrent read access without blocking other consumers.  
For cross-host communication (e.g., between scheduler and remote workers), each host maintains a BIC replica, and a background daemon handles data synchronization across hosts.

For combine—aggregating sampling results from all samplers back to the scheduler—we design \emph{BIC-O}, a TCP-based multi-producer ring buffer. Each slot in BIC-O contains multiple subslots, one per sampler. In iteration $n$, sampler $i$ writes its output to subslot $i$ of slot $n \bmod N$. Once all subslots are filled, the scheduler detects that the microbatch is completed. Since each sampling result is typically just a token ID (i.e., an integer), this TCP-based approach is lightweight and avoids the complexity of emulating shared memory across hosts.

\section{Evaluation}
\label{sec:evaluation}

This section answers four key questions:  

\noindent
(1) Does \sysname deliver performance gains across diverse parallel inference configurations? (\S\ref{sec:tput_gain})

\noindent
(2) How does \sysname impact end-to-end inference latency across diverse configurations? (\S\ref{sec:latency_gain})

\noindent
(3) How does \sysname affect compute resource utilization across diverse configurations? (\S\ref{sec:utilization})

\noindent
(4) What is the contribution of each component in \sysname's design? (\S\ref{sec:ablation})

\subsection{Experimental Setup}
\label{sec:exp_setup}
\textbf{Testbed.} Our evaluation is conducted on following testbeds: {\tt H100-NVL}, {\tt A100-NVL}, and {\tt A100-PCIe}. Each server in H100-NVL and A100-NVL is equipped with 8 H100 or A100 GPUs connected via NVLink, respectively, while each A100-PCIe server contains 8 A100 GPUs connected via PCIe. All GPUs have 80\,GB memory. Each server also features a 192-logical-core Intel(R) Xeon(R) Platinum 8468 CPU and 2\,TB of host memory. Servers within the same testbed are connected via 400\,Gbps ConnectX-7 NICs.

\noindent
\textbf{Baseline.}
We compare \sysname against two of the most popular inference engines: {\tt vLLM}~\cite{kwon2023vllm} and {\tt SGLang}~\cite{zheng2024sglang}. Since SGLang does not support PP, we evaluate only its TP mode. For vLLM, we evaluate both TP and PP modes.

\begin{figure*}
    \centering
    \setlength{\abovecaptionskip}{-1pt}
    \includegraphics[width=\textwidth]{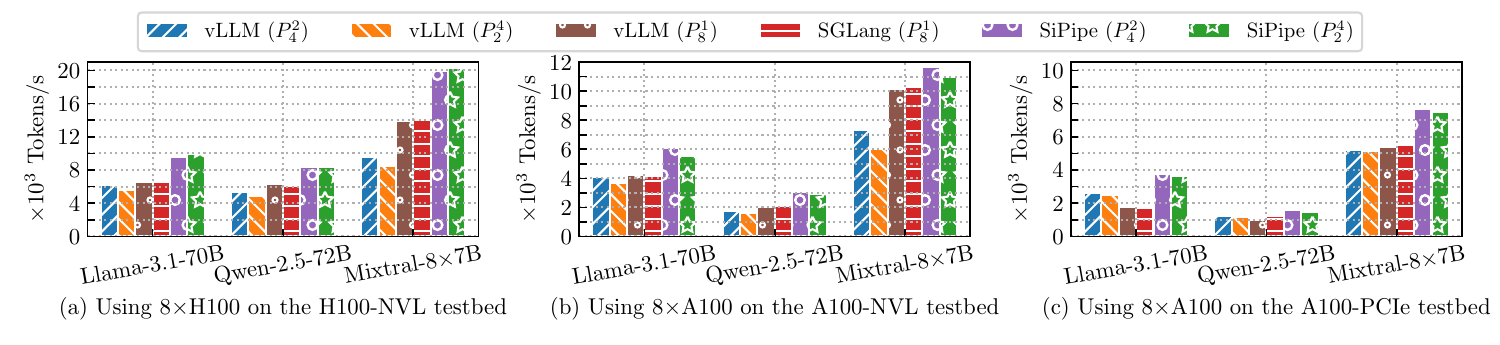}
    \caption{Throughput comparison of different engines under various parallel configurations, evaluated on multiple testbeds. Each configuration is denoted as $P^{i}_{j}$, where PP degree $p = i$ and the TP degree $t = j$. Specifically, $P^1_t$ denotes the pure TP mode.}
    \label{fig:bench_tput_08GPU}
    \crunch
    \crunch
    \crunch
\end{figure*}

\noindent
\textbf{Models.} We evaluate \sysname using six LLMs, categorized into two groups: \emph{large} LLMs—Llama-3.1-70B~\cite{llama-3.1}, Qwen-2.5-72B~\cite{qwen2.5}, and Mixtral-8$\times$7B~\cite{jiang2024mixtral}; and \emph{ultra-large} LLMs —DeepSeek V2.5~\cite{liu2024deepseek}, DeepSeek V3~\cite{liu2024deepseekv3}, and Llama-3.1-405B~\cite{llama-3.1}. Based on deployment experience, \emph{large} LLMs are typically served on 8 GPUs~\cite{llama2deploy, llama2followup}, whereas the \emph{ultra-large} models require 16 GPUs because their weights alone saturate the memory of 8 GPUs. 
We focus on LLMs of this scale because PP is primarily intended for multi-GPU deployments. In contrast, for smaller LLMs (e.g., 14B), pure TP is a more suitable choice. Optimizing TP performance for small LLMs is beyond the scope of this work. Data parallelism is also omitted: instances with fewer GPUs lack sufficient memory for both model weights and KV caches.

\noindent
\textbf{Configuration.}
The default batch size is 512 for 8-GPU and 1024 for 16-GPU experiments. In pure TP mode, PP is disabled ($p = 1$), and the TP degree $t$ equals the total number of GPUs. In PP mode, each stage forms a TP group assigned to a contiguous block of LLM layers, with $pt$ equal to the GPU count. Empirically, setting $t = 4$ per stage delivers near-optimal intra-stage efficiency~\cite{vllm_tunning}. 
Samplers are colocated with the final stage. On this node, each worker occupies an isolated CPU core for \emph{input preparation} (8 cores total), while the remaining cores are dedicated to \emph{sampling}.

\noindent
\textbf{Workload.} To ensure fair and reproducible comparison, we randomly select a fixed set of prompts from the ShareGPT dataset~\cite{shareGPT} and reuse them across all experiments. Decoding is performed using all common sampling strategies, including top-$p$, top-$k$, min-$p$, temperature, and penalties for repetition, presence, and frequency.

\noindent
\textbf{Metrics.} We report throughput, per-token latency, and resource utilization. Throughput is measured in tokens per second, and per-token latency reflects the average generation time per token. Utilization metrics reflect hardware efficiency. 

\subsection{Throughput Gains under Diverse Parallel Settings}
\label{sec:tput_gain}

\noindent
\textbf{Throughput across LLMs and testbeds.}
Figure~\ref{fig:bench_tput_08GPU}(a) shows the throughput of \emph{large} LLMs on the {\tt H100-NVL} testbed using different inference engines and configurations. Compared to vLLM ($P^{2}_{4}$), \sysname achieves a performance gain of $1.6\times$ to $2.1\times$ by eliminating pipeline bubbles that degrade vLLM’s PP mode, especially as the number of stages increases (e.g., from $P^{2}_{4}$ to $P^{4}_{2}$ in vLLM). In this single-node scenario, we also observe that both vLLM and SGLang in pure TP mode outperform vLLM’s PP mode due to NVLink-enabled efficient GPU communication and better TP scalability. However, as each PP stage has a smaller TP degree ($2$ or $4$) compared to the pure TP modes ($t=8$), using PP cuts \emph{all-reduce} overhead~\cite{shoeybi2019megatron, bavarian2022efficient} and is inherently more efficient. Building on this, \sysname further addresses the inefficiencies of inter-stage coordination in prior PP implementation, enabling it to outperform pure TP modes in overall throughput.

\begin{figure}
    \centering
    \setlength{\abovecaptionskip}{-1pt}
    \includegraphics[width=\columnwidth]{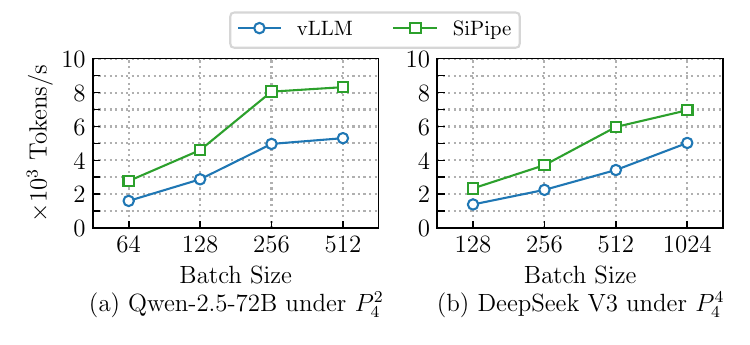}
    \caption{Effect of batch size on inference throughput.}
    \label{fig:tput_batch}
    \crunch
    \crunch
    \crunch
\end{figure}

Figure~\ref{fig:bench_tput_08GPU}(b) and Figure~\ref{fig:bench_tput_08GPU}(c) show results on the {\tt A100-NVL} and {\tt A100-PCIe} testbeds, respectively. \sysname consistently outperforms existing frameworks across GPU architectures and interconnects, demonstrating strong generality and robustness in both NVLink- and PCIe-based environments.
Notably, the speedup on A100 is smaller than on H100, as expected. H100's $3\times$ higher compute capability makes pipeline bubbles a more significant bottleneck relative to the \emph{forward} pass. On the A100, by contrast, the \emph{forward} pass dominates the runtime. This highlights an important trend—future accelerators will further reduce the relative cost of \emph{forward}, making pipeline inefficiencies more pronounced. \sysname aligns with this trend by minimizing stalls and maximizing pipeline efficiency.

Next, we revisit the 16-GPU inference results for \emph{ultra-large} LLMs, previously shown in Figure~\ref{fig:bench_tput_16GPU_H100}. In this cross-node setting, \sysname consistently outperforms all other inference engines, achieving $1.4\times$-$1.7\times$ improvement over the best baseline. The speedup is relatively lower than that observed for \emph{large} LLMs, primarily because the \emph{forward} pass in \emph{ultra-large} models takes longer to compute, which reduces the relative impact of pipeline bubbles. Notably, pure TP configurations perform worse than PP in this scenario. As confirmed in prior studies~\cite{shoeybi2019megatron}, the \emph{allreduce} communication in TP scales poorly across machines, making them a major performance bottleneck.

\noindent
\textbf{Impact of batch size on throughput.}
Because production workloads rarely use a fixed batch size, we swept the batch size to gauge \sysnames robustness and scalability. Figure~\ref{fig:tput_batch} presents throughput comparison for both \emph{large} and \emph{ultra-large} LLMs, each with a representative LLM. Under these batch sizes, \sysname consistently delivers stable performance gains. Compared to vLLM, \sysname achieves a speedup of 1.6-1.7$\times$ on Qwen-2.5-72B and 1.4-.17$\times$ on DeepSeek V3, demonstrating that the optimizations in \sysname remain effective even under low-load scenarios (i.e., small batch sizes). 

\begin{figure}
    \centering
    \setlength{\abovecaptionskip}{-1pt}
    \includegraphics[width=\columnwidth]{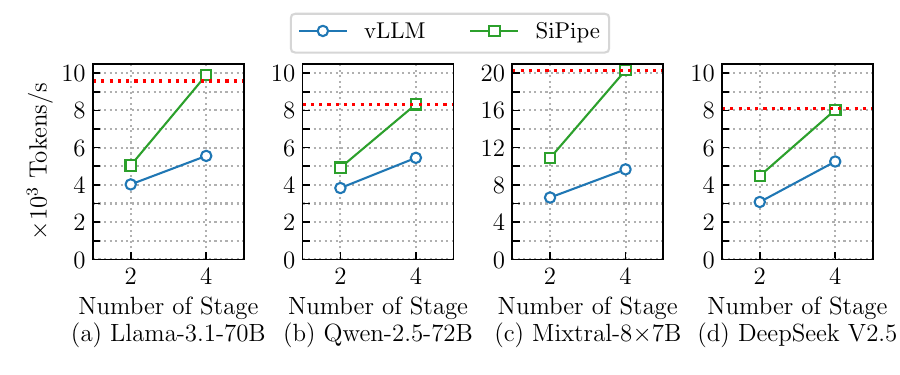}
    \caption{Effect of the number of pipeline stages (i.e., PP degree $p$) on throughput. In the first three subfigures, each stage uses a TP degree of $t{=}2$, while the last subfigure uses $t{=}4$. The dashed line denotes the performance of \sysname under the default configuration.}
    \label{fig:tput_scalability}
    \crunch
    \crunch
    \crunch
\end{figure}
\noindent
\textbf{Scalability under multi-GPU settings.}
Figure~\ref{fig:tput_scalability} demonstrates the scalability of \sysname with varying number of GPUs. In LLM inference, it is typical to reserve KV cache memory amounting to 2--3$\times$ the LLM size~\cite{shoeybi2019megatron, kwon2023vllm}. For example, deploying a 70B LLM generally requires $>4$ GPUs~\cite{llama2followup, vllm_tunning}. However, due to poor scalability in existing inference engines, adding GPUs to a single instance often yields less benefit than running multiple smaller ones--an approach that only works when all sequences are short, which is rarely the case in production. 

We therefore compare one-instance scaling directly.
\emph{large} LLMs use $t=2$ while  \emph{ultra-large} models use $t=4$.
As shown in Figure~\ref{fig:tput_scalability}, vLLM consistently falls short of linear scalability, achieving only $1.4\times$ to $1.7\times$ speedup. In contrast, \sysname achieves $1.8\times$ to $2.0\times$ scaling. The improvement arises because doubling GPUs expands KV-cache capacity super-linearly, relieving memory contention; \sysnames pipeline optimizations convert this headroom into higher effective throughput.
Improved scalability lets more memory be devoted to KV caches, cutting request preemption and expensive CPU–GPU cache transfers in large workloads. It also reduces the number of instances required to saturate the cluster, easing inter-instance load imbalance \cite{zheng2024sglang}.

\begin{figure}[t]
    \centering
    \setlength{\abovecaptionskip}{-1pt}
    \includegraphics[width=\linewidth]{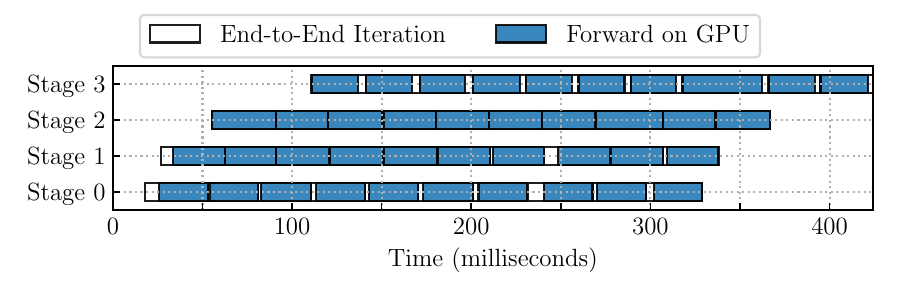}
    \caption{Per-iteration execution breakdown of each pipeline stage using \sysname ($P^4_4$) on 16 H100 GPUs in the H100-NVL testbed with the DeepSeek V3.}
    \label{fig:breakdown_iteration_deepseek_v3_sillm}
    \crunch
    \crunch
    \crunch
    \crunch
    \crunch
    \crunch
\end{figure}
\noindent
\textbf{Pipeline efficiency and bubble analysis.}
We analyze the execution timeline to evaluate the pipeline efficiency of \sysname. Figure~\ref{fig:breakdown_iteration_deepseek_v3_sillm} presents a per-iteration execution breakdown of each stage using DeepSeek V3—the largest LLM in our evaluation and the one with the lowest relative performance gain. As shown, most pipeline bubbles have been eliminated, and workload is reasonably balanced across stages. The last stage is no longer a performance bottleneck. Some bubbles remain in the pipeline due to intrinsic workload asymmetry arising from the MoE architecture, which is orthogonal to the pipeline optimizations proposed in this paper. 

\subsection{Latency Reduction under Different Configurations}
\label{sec:latency_gain}

\noindent
\textbf{Time-to-First-Token (TTFT).} TTFT measures the delay from request arrival to the first output token and is mainly affected by scheduling and queuing, which are outside the scope of \sysname.

\begin{figure}
    \centering
    \setlength{\abovecaptionskip}{-1pt}
    \includegraphics[width=\columnwidth]{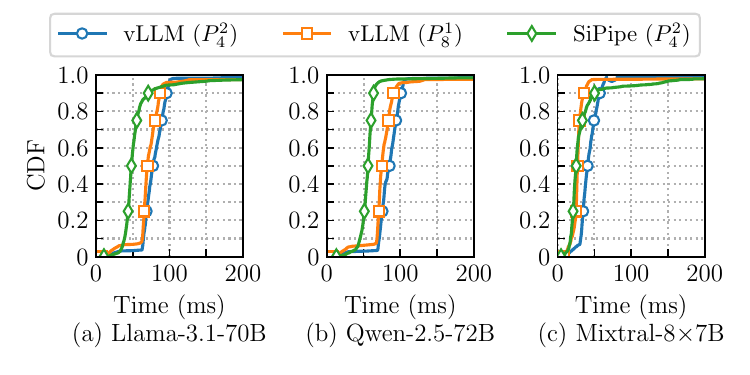}
    \caption{TPOT comparison of different enginess, evaluated with 8$\times$H100 on H100-NVL testbed.}
    \label{fig:bench_tpot_08h100}
    \crunch
    \crunch
    \crunch
    \crunch
    \crunch
    \crunch
\end{figure}
\begin{figure}
    \centering
    \setlength{\abovecaptionskip}{-1pt}
    \includegraphics[width=\columnwidth]{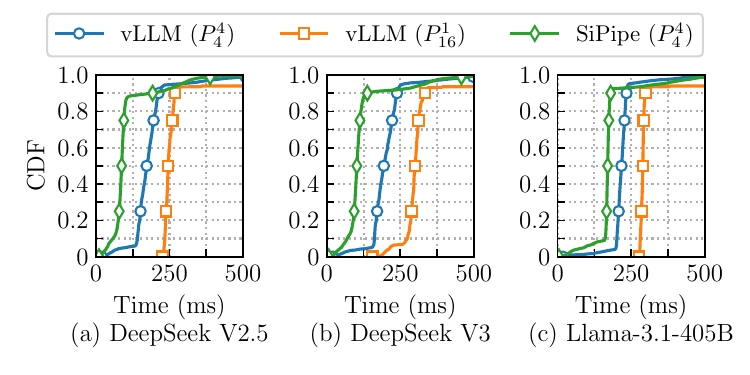}
    \caption{TPOT comparison of different enginess, evaluated with 16$\times$H100 on H100-NVL testbed.}
    \label{fig:bench_tpot_16h100}
    \crunch
    \crunch
    \crunch
    \crunch
    \crunch
    \crunch
\end{figure}

\noindent
\textbf{Time-per-Output-Token (TPOT).} 
TPOT measures the average latency incurred for generating each output token during the decoding phase. It directly reflects the efficiency of per-iteration execution in the pipeline. Figure~\ref{fig:bench_tpot_08h100} and Figure~\ref{fig:bench_tpot_16h100} present the cumulative distribution (CDF) of per-iteration TPOT across \emph{large} and \emph{ultra-large} LLMs, respectively. For \emph{large} LLMs, \sysname reduces the average TPOT by up to 31\%, while for \emph{ultra-large} LLMs the reduction reaches 43\%. These improvements stem from \sysnames elimination of pipeline bubbles, which significantly shortens the execution time of each iteration.

We observe more substantial gains in \emph{ultra-large} models, where deeper pipelines amplify the impact of PP bubbles. In PP with degree $p$, a stall of duration $\Delta t$ in any single stage can delay the entire pipeline by up to $p \Delta t$, as all $p$ stages are sequentially blocked. Notably, for Mixtral-8$\times$7B, \sysnames TPOT is comparable to that of the pure TP baseline. This model features faster \emph{forward} execution than others in its category due to its sparse MoE architecture. In single-node settings, pure TP benefits from avoiding inter-stage communication, while \sysname's design introduces coordination overhead that offsets its efficiency gains at the TPOT level.


\subsection{Resource Utilization Analysis}
\label{sec:utilization}
\begin{figure}[t]
    \centering
    \setlength{\abovecaptionskip}{-1pt}
    \includegraphics[width=\linewidth]{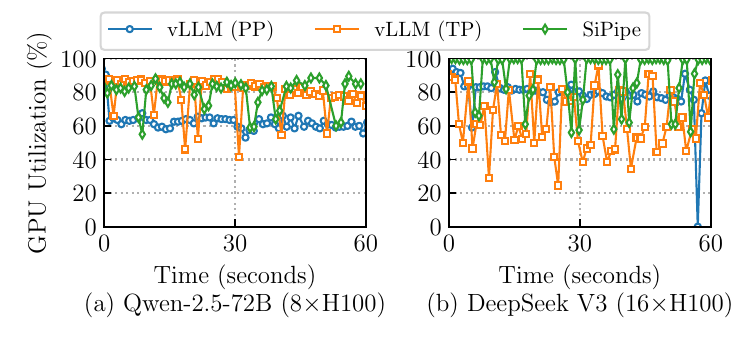}
    \caption{GPU utilization over time on H100-NVL testbed. vLLM (TP) and \sysname adopt the default configuration, while vLLM (TP) using pure TP configuration.}
    \label{fig:bench_util_gpu}
    \crunch
    \crunch
    \crunch
\end{figure}

\noindent
\textbf{GPU utilization.} We report GPU utilization to assess how effectively each approach leverages hardware resources. Figure~\ref{fig:bench_util_gpu} presents the average GPU utilization across all devices for both \emph{large} and \emph{ultra-large} LLMs.

In the 8-GPU \emph{large} model setting (Figure~\ref{fig:bench_util_gpu}(a)), \sysname achieves an average utilization of approximately 85\%, 23\% higher than that of vLLM (PP). While \sysname and the pure TP baseline exhibit similar utilization, pure TP consumes more streaming multiprocessor (SM) resources for \emph{all-reduce} communication. Consequently, \sysname delivers higher throughput under comparable GPU utilization.

In the 16-GPU \emph{ultra-large} setting (Figure~\ref{fig:bench_util_gpu}(b)), \sysname achieves over 95\% average utilization, surpassing all baselines. This improvement stems from eliminating pipeline bubbles in PP baselines and avoiding the underutilization caused by slow cross-node \emph{all-reduce} in pure TP.

\begin{figure}[t]
    \centering
    \setlength{\abovecaptionskip}{-1pt}
    \includegraphics[width=\linewidth]{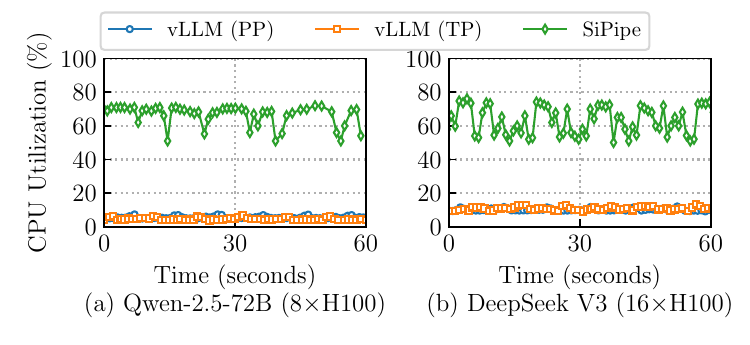}
    \caption{CPU utilization over time on H100-NVL testbed. vLLM (TP) and \sysname adopt the default configuration, while vLLM (TP) using pure TP configuration.}
    \label{fig:bench_util_cpu}
    \crunch
    \crunch
    \crunch
\end{figure}

\noindent
\textbf{CPU utilization.} 
A key distinction of \sysname compared to other inference engines is its effective use of otherwise underutilized CPU resources to accelerate inference. Figure~\ref{fig:bench_util_cpu} \sysname shows that this strategy lifts average CPU utilization by approximately 60\%, confirming that the CPU now contributes meaningfully across the inference process.

We also observe differences in CPU utilization patterns across LLMs. In Figure~\ref{fig:bench_util_cpu}(a), \sysname exhibits higher and more stable CPU usage compared to Figure~\ref{fig:bench_util_cpu}(b). While the sampling workload per iteration is similar across LLMs, the fraction of time spent on \emph{forward}  varies significantly —\emph{ultra-large} LLMs have longer \emph{forward}. Since sampling tasks are executed after \emph{forward}, faster \emph{forward} leads to more frequent sampling tasks, driving higher and steadier CPU utilization. Conversely, the longer \emph{forward} in \emph{ultra-large} LLMs result in more bursty CPU utilization patterns. 

\subsection{Ablation Studies}
\label{sec:ablation}

\begin{figure}
    \centering
    \setlength{\abovecaptionskip}{-1pt}
    \includegraphics[width=\columnwidth]{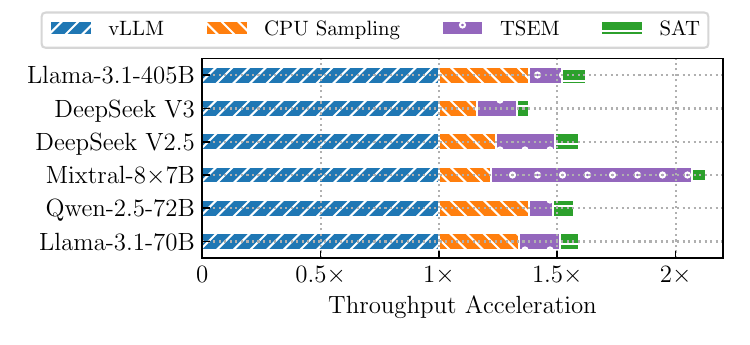}
    \caption{Incremental performance improvements contributed by each individual design.}
    \label{fig:ablation}
    \crunch
    \crunch
    \crunch
\end{figure}

Figure~\ref{fig:ablation} presents the performance improvements contributed by each design component of \sysname under the default configuration on the H100-NVL testbed. The results show that most gains come from CPU sampling ($16\%\text{--}38\%$) and TSEM ($80\%$ for Mixtral, as its fast \emph{forward} makes input preparation more dominant, and $10\%\text{--}24\%$ for others), consistent with the per-iteration breakdown analysis in Section~\ref{sec:observation}, which identifies \emph{intra-stage bubbles} and \emph{load-imbalance bubbles} as the main overheads in pipeline parallelism. Additionally, SAT contributes a $3\%\text{--}6\%$ improvement by eliminating stalls that arise when an upstream stage finishes its forward pass but must wait for the downstream stage to become ready to receive hidden states. 

\section{Related Work}
Besides the classic implementation of parallelism in open-source inference engines~\cite{zheng2024sglang, kwon2023vllm},
recent developments in PP for LLM inference reveal a dynamic shift from simple layer-based designs toward rich, multi-dimensional pipelined architectures that adapt to model structure and deployment scenarios. Classic vertical pipelining— partitioning consecutive layers across devices—has evolved through micro-batching into more nuanced systems that support sequence slicing (PipeLLM~\cite{tan2025pipellm}, HPipe~\cite{ma2024hpipe}), patch-level parallelism for vision models (PipeFusion~\cite{wang2024pipefusion}), and token-level speculative execution (PipeInfer~\cite{butler2024pipeinfer}), all tailored to minimize pipeline bubbles and boost utilization. Parallel efforts such as PISeL~\cite{rahimijafari2024pisel}, PLANCK~\cite{lin2024planck}, Quart~\cite{lin2024quart}, and gLLM~\cite{guo2025gllm} underscore the importance of latency-awareness, cold-start avoidance, and SLO-aligned scheduling, signaling an industry-wide pivot toward dynamic, runtime-sensitive pipeline orchestration.

Furthermore, MoE models are catalyzing a new frontier of pipeline design by leveraging expert-disaggregated pipelining. 
Systems like MegaScale–Infer~\cite{zhu2025megascaleinfer} and Pipeline MoE~\cite{chen2023pipeline_moe} decouple attention and FFN modules or experts, orchestrating ping-pong microbatches to overlap computation and communication efficiently and reflexively across heterogeneous GPUs—achieving significant throughput and cost gains. \sysname can be compatible with these techniques as they parallel and accelerate the inference in different dimensions while meeting the same targets of strict latency and utilization.

\section{Conclusion}

This paper identifies key scalability bottlenecks in PP for LLM inference, including load-imbalance bubbles, intra-stage bubbles, and inter-stage bubbles. To overcome these challenges, we leverage the underutilized CPU resources on the host by offloading the sampling process to the CPU, introducing an asynchronous CPU executor, and designing a CPU-assisted communication module. These optimizations collectively yield up to $2\times$ improvement in GPU inference throughput on large-scale LLMs. Our findings demonstrate that carefully orchestrating CPU-GPU collaboration is critical for unlocking the full potential of PP inference.

\bibliographystyle{ACM-Reference-Format}
\bibliography{acmart}

\appendix

\section{Performance Modeling}
\label{appendix}

\begin{table}[t!]
    \caption{Notations and their meanings.}
    \label{tab:notation1}
    \centering
    \small
    \begin{tabular}{|c|l|}
        \hline
        \textbf{Notation}    & \multicolumn{1}{c|}{\textbf{Meaning}}        \\ \hline
        $L$         & Number of transformer layers                          \\ \hline
        $C$         & Per-layer computation time of models                  \\ \hline
        $N$         & Total number of GPUs                                  \\ \hline
        $p$         & Pipeline parallelism degree                           \\ \hline
        $t$         & Tensor parallelism degree                             \\ \hline
        $\alpha$    & Launch delay of communication                         \\ \hline
        $s$         & Sequence length                                       \\ \hline
        $b$         & Batch size                                            \\ \hline
        $h$         & Hidden layer size                                     \\ \hline
        $B_1$       & Intra-node bandwidth                                  \\ \hline
        $B_2$       & Inter-node bandwidth                                  \\ \hline
        $m$         & Number of microbatches                                \\ \hline
        $n$         & Number of hosts                                       \\ \hline
    \end{tabular}%
\end{table}

This section derives analytic expressions for throughput and per-token delay in model-parallel LLM inference deployments. All symbols are defined in Table~\ref{tab:notation1}. We first analyze a single-node configuration, where a single inference engine is partitioned across multiple GPUs using either tensor parallelism (TP) or pipeline parallelism (PP), and then extend the LLM to cross-node settings.


\subsection{Pure Tensor Parallelism}
Each transformer layer performs two \emph{allreduce} communication operations under TP. Including start-up latency, the communication cost on $N$ GPUs is
\begin{equation}
    t_{\text{comm-TP}}(N)=2L \times (\alpha\log_2N+2sbh/B_1).
\end{equation}
Therefore, the overall throughput of the inference system is:
\begin{equation}
\begin{aligned}
    T(N) &= \frac{b}{t_{\text{comp-TP}}(N)+t_{\text{comm-TP}}(N)} \\
         &= \frac{b}{\frac{L\,C}{N}+2L\alpha\log_2N+4sbhL/B_1}.
\end{aligned}
\end{equation}
Correspondingly, the per-token delay—the reciprocal of throughput—is:
\begin{equation}
    D(N) = \frac{L\,C}{N}+2L\alpha\log_2N+4sbhL/B_1.
\end{equation}

\subsection{Pure Pipeline Parallelism}
Pipeline parallelism splits a batch of $b$ sequences into $m$ microbatches , so each stage produces $b/m$ tokens per iteration.
For one stage the latency is:
\begin{equation}
    t_{\text{stage}}=\frac{L\,C}{N}+sbh/B_1.
\end{equation}
Hence overall throughput is:
\begin{equation}
\begin{aligned}
    T(N) &= \frac{b/m}{t_{\text{stage}}} \\
         &= \frac{b/m}{\frac{L\,C}{N}+sbh/B_1}.
\end{aligned}    
\end{equation}

A single token experiences $N$ compute phases and $N-1$ inter-stage transfers, so its end-to-end delay is
\begin{equation}
    D(N)=N \times \frac{L\,C}{N}+(N-1) \times sbh/B_1.
\end{equation}

\subsection{Hybrid Model Parallelism}
In production, tensor and pipeline parallelism are often enabled together to fit larger models, so the system incurs communication overheads from each.
Let $t$ be the tensor-parallel degree and $p$ the pipeline-parallel degree ($p \times t = N$).
Combining the two previous models yields:
\begin{equation}
\begin{aligned}
    T(p,t) &= \frac{b/m}{\frac{L\,C}{N}+sbh/B_1+\frac{2L}{p}(\alpha\log_2t+2sbh/B_1)}; \\
    D(p,t) &= p (\frac{L\,C}{N}+\frac{2L}{p}(\alpha\log_2t+2sbh/B_1)) \\
    &+(p-1) sbh/B_1.
\end{aligned}
\end{equation}

\subsection{Multi-node Model Parallelism}
When an inference instance spans multiple nodes, part of the communication occurs over a slower inter-node network with bandwidth $B_2$. Assuming that $n$ hosts are connected through this network, then the performance of pure tensor parallelism is:
\begin{equation}
\begin{aligned}
    T(N) = \frac{b}{\frac{L\,C}{N}+2L(\alpha\log_2N+2sbh/B_2)}, \\
    D(N) = \frac{L\,C}{N}+2L(\alpha\log_2N+2sbh/B_2).
\end{aligned}
\end{equation}

For pure pipeline parallelism, only the $n-1$ inter-stage transfers that cross hosts use bandwidth $B_2$; the remaining $N-n$ transfers stay on the faster intra-node link $B_1$:
\begin{equation}
\begin{aligned}
    T(N) &= \frac{b/m}{\frac{L\,C}{N}+sbh/B_2}, \\
    D(N) &= L\,C+(n-1) sbh/B_2+(N-n) sbh/B_1.
\end{aligned}
\end{equation}

Put them together, the combined throughput and per-token delay are:
\begin{equation}
\begin{aligned}
    T(p,t) &= \frac{b/m}{\frac{L\,C}{N}+sbh/B_2+2L(\alpha\log_2N+2sbh/B_2)}, \\
    D(p,t) &= L\,C+(n-1) sbh/B_2+(N-n) sbh/B_1 \\ 
    &+2L(\alpha\log_2N+2sbh/B_2).
\end{aligned}
\end{equation}

From the models above we note two insights:

(1) \textbf{Throughput scaling.} Pipeline parallelism scales almost linearly with the number of GPUs, whereas tensor parallelism hits a throughput ceiling: although computation is divided further, the communication volume stays constant (or even grows), so GPU utilization plateaus.

(2) \textbf{Latency trends.} Increasing the tensor-parallel degree quickly lowers per-token latency, while deeper pipeline parallelism lengthens it because each token must traverse more stages.

\end{document}